\documentclass[prd,amsmath,notitlepage,twocolumn,10pt,nofootinbib]{revtex4-2}

\usepackage{graphicx}
\usepackage{float}
\usepackage{epstopdf,cancel}
\usepackage{epsf}
\usepackage[caption=false]{subfig}
\usepackage{latexsym,bbm,euscript}
\usepackage{amssymb,amsmath}
\usepackage{mathtools} % also for \coloneqq and := via \vcentcolon
\usepackage{mathrsfs}
\usepackage{tensor}
\usepackage{braket}

\usepackage{times,graphics}
\usepackage{soul,xcolor}
\usepackage[normalem]{ulem}
\usepackage{setspace}

\usepackage{scalerel}
\usepackage{tikz}
\usetikzlibrary{svg.path}
\definecolor{orcidlogocol}{HTML}{A6CE39}
\tikzset{
  orcidlogo/.pic={
    \fill[orcidlogocol] svg{M256,128c0,70.7-57.3,128-128,128C57.3,256,0,198.7,0,128C0,57.3,57.3,0,128,0C198.7,0,256,57.3,256,128z};
    \fill[white] svg{M86.3,186.2H70.9V79.1h15.4v48.4V186.2z}
    svg{M108.9,79.1h41.6c39.6,0,57,28.3,57,53.6c0,27.5-21.5,53.6-56.8,53.6h-41.8V79.1z M124.3,172.4h24.5c34.9,0,42.9-26.5,42.9-39.7c0-21.5-13.7-39.7-43.7-39.7h-23.7V172.4z}
    svg{M88.7,56.8c0,5.5-4.5,10.1-10.1,10.1c-5.6,0-10.1-4.6-10.1-10.1c0-5.6,4.5-10.1,10.1-10.1C84.2,46.7,88.7,51.3,88.7,56.8z};
  }
}
\newcommand\orcidlink[1]{\href{https://orcid.org/#1}{\mbox{\scalerel*{
  \begin{tikzpicture}[yscale=-1,transform shape]
    \pic{orcidlogo};
  \end{tikzpicture}}{X}}}}

\usepackage{url,hyperref}
\hypersetup{colorlinks,linkcolor={blue!55!black},citecolor={red!45!black},urlcolor={blue!45!black},breaklinks=true}

\newcommand{\pbh}{$\Phi$BH}

\def\sg{\textsl{g}}
\def\6{{\langle}}
\def\9{{\rangle}}
\def\pad{\partial}
\def\half{\tfrac{1}{2}}

\def\cA{\mathcal{A}}
\def\cO{\mathcal{O}}

\def\mbh{\mathrm{MBH}}
\def\ah{\mathrm{H}}
\def\PG{\mathrm{PG}}

\def\bt{{\bar{t}}}

\newcommand{\defeq}{{\vcentcolon=}}
\newcommand{\eqdef}{{=\vcentcolon}}

\newcommand{\be}{\begin{equation}}
\newcommand{\ee}{\end{equation}}

\begin{document}
	
	\title{ Apparent horizon as a membrane}

	\author{Daniel R. Terno\,\orcidlink{0000-0002-0779-0100}}
	\email{daniel.terno@mq.edu.au}
	
	\affiliation{School of Mathematical and Physical Sciences, Macquarie University, NSW 2109, Australia}

\begin{abstract}
%Near-horizon geometry of physical black holes   that form within a finite time for distant observers differs from that of eternal black holes. While in spherical symmetry the identification of possible dynamical %solutions is complete and the qualitative differences are remarkable, it remains unclear whether these lead to potentially observable signature. We construct an approximate near-horizon metric that encapsulates the %differences and is suitble for use in modelling.

The requirement that a trapped spacetime domain forms in finite time for distant observers is logically possible and sometimes unavoidable, but its consequences are not yet fully understood. In spherical symmetry, the characterization of the near-horizon geometry of these physical black holes is complete and shows marked differences from their eternal counterparts. Whether these differences lead to observable signatures remains unclear. We construct an approximate near-horizon metric that encapsulates them and is suitable for modeling.
The timelike apparent horizon of physical black holes provides a natural surface for a consistent membrane description: we obtain closed-form expressions for the redshift, proper acceleration, and extrinsic curvature, and assign a two-dimensional viscous-fluid stress tensor via junction conditions. These results also provide an additional perspective on the relation between Rindler and near-horizon geometries. Among dynamical generalizations of surface gravity, only a subset applies to these models. We complete their analysis and recover the intuitive definition of surface gravity --- the acceleration in the frame of a near-horizon observer, redshifted to infinity --- directly from the membrane acceleration.

\end{abstract}

	\maketitle

\section{Introduction}
 More than a hundred astrophysical black holes --- dark, massive, ultra-compact objects --- have been identified \cite{LIGO:23,EHT:22}. While all the observations so far are consistent with the classical Kerr solution to Einstein--Hilbert gravity, there are more than a dozen classes of models that aim to provide an alternative description of the observed objects \cite{CP:19,BCC:25}. Their proliferation is possible because the defining textbook feature of a black hole --- the event horizon \cite{HE:73,FN:98,F:15,AK:25} --- is a teleological concept and is, in principle, inaccessible to local observers \cite{V:14,B:17}.

In classical general relativity, \textit{mathematical} black holes (MBHs) --- spacetime domains whose interior is causally disconnected from the exterior by a null event horizon --- arise only as the asymptotic end state of gravitational collapse. Accepting their formation in finite time entails serious mathematical difficulties and potential paradoxes \cite{FN:98,BD:84,BMPS:95,B:17,MMT:22}.

In contrast with an MBH, a \textit{physical} black hole  \cite{F:14b} (\pbh)\footnote{The term “physical black hole” was introduced in Ref.~\cite{F:14b} and is abbreviated as PBH in Ref.~\cite{MMT:22} and elsewhere. Because PBH is also the standard acronym for “primordial black hole,” we adopt a distinct abbreviation here to avoid confusion.} is a light-trapping spacetime domain. A locally defined apparent horizon, which is its boundary, is what is actually determined in numerical simulations, particularly in dynamical scenarios \cite{RZ:13}. In general, it depends on the foliation of spacetime, which is observer-dependent. We additionally require it to form within a finite time as measured by a distant observer \cite{MMT:22,M:23}. As such,it belongs
to the general dynamical-horizon framework \cite{AK:25}, but constitute a special case with
distinct geometric and physical properties.

From the perspective of a distant observer, collapse beyond the Buchdahl \cite{B:59,ST:25} limit can proceed in one of three ways: (i) perpetual ongoing collapse, where a horizon exists only as an asymptotic ($t \rightarrow \infty$) concept and, for any $t<\infty$, the nearly frozen configuration remains horizonless by definition; (ii) formation of a horizonless ultra-compact object, either at some finite time $t_{\min}$ or asymptotically as $t \rightarrow \infty$; (iii) formation of an apparent horizon in finite time $t_{\mathrm{f}} < \infty$, as measured by the clock of a distant observer.

Of the three, (i) and (iii) lead to black holes in different senses: (i) approaches an MBH asymptotically ($t \rightarrow \infty$), whereas (iii) yields a black hole in finite time. The Kerr and Schwarzschild MBHs are the asymptotic limit of (i). If we are interested in black holes whose key features have formed by now and are, in principle, observable, we have to select (iii).

At present, most efforts to identify the true nature of astrophysical black holes (ABHs) focus on comparing the models of the first two scenarios \cite{CP:19,BCC:25,FLV:24}. Event horizons are not directly observable, but all current ABH data are consistent with models that explicitly include them. Consequently. viable horizonless proposals have only small, model-dependent deviations that upcoming searches may detect or exclude.

Introducing \pbh s raises two questions. First, do MBHs and \pbh s differ in geometric or physical properties? Second, if so, do those differences yield distinct predictions for distant local observers? The first is answered in the affirmative \cite{MMT:22}, and there are suggestive indications supporting the second \cite{DMSS:24}. The main goal of this work is to develop an observation-facing framework that translates these differences into testable signatures.

The requirements of finite $t_{\mathrm{f}}$ and minimal regularity conditions of the apparent horizon allow, in spherical symmetry, a complete classification of the near-horizon geometries. Some of their properties are conceptually  different from both the Schwarzschild solution and regular black hole models. In Section~\ref{s:prop}, after reviewing our self-consistent framework and the essential properties of the $\Phi$BH solutions, we discuss how the dynamical solutions can describe astrophysically relevant nearly static configurations.

The apparent horizon of a \pbh\ is a timelike surface. This makes it natural to apply the membrane (or stretched-horizon) formalism. In the standard membrane paradigm \cite{TPM:86}, the membrane is introduced as an auxiliary timelike surface placed at a small coordinate distance outside a null event horizon. In the present setting, it can be identified with a geometrically defined hypersurface and thus requires no additional prescription. After briefly introducing the paradigm in Section~\ref{s:memb}, we discuss the consequences of this identification. We also examine aspects of the relation between the near-horizon geometry of a \pbh\ and the Rindler geometry. The membrane viewpoint provides a useful framework for analysing generalisations of surface gravity to dynamical spacetimes, which we report in Section~\ref{s:kappa}. We conclude with the discussion and an outline of future work in Section~\ref{s:dis}.

Throughout this article, we work in natural units $G=c=\hbar=1$. The outer apparent horizon is at the Schwarzschild radius $r_\sg(t)=r_+(v)$, that we express using the time coordinate $t$ and the advanced null coordinate $v$, respectively. When convenient, the coordinate distance from the Schwarzschild radius is denoted $x\defeq r-r_\sg$ or $y\defeq r-r_+$, depending on the coordinates used. A prime denotes differentiation: $r_+'\defeq dr_+/dv$ and $r'_\sg\defeq dr_\sg/dt$. To shorten descriptions, we refer to a distant static observer as Bob, a horizon-crossing observer as Alice, and a static or comoving observer as Eve. We use fractional subscripts only for small fractions and omit the fraction bar, so $h_{1/2}$ is written as $h_{12}$.

\section{Properties of physical black holes}\label{s:prop}
\subsection{The framework }

The very posing of the collapse trilemma assumes
the classical geometric picture inherent to semiclassical gravity. The phenomena such as generation of gravitational waves and their propagation, motion of test particles, or light propagation  are assumed to occur as in  general relativity or some alternative classical theory of gravity.  The semiclassical description incorporates quantum expectation values for the renormalised energy-momentum tensor (EMT) of matter fields into the classical framework.   In this setting, the metric $\sg_{\mu\nu}$  is a solution to the  semiclassical Einstein equations \cite{BD:84,FW:96,K:07,HV:20}
	\be
R_{\mu\nu}-\tfrac{1}{2} \sg_{\mu\nu}R=8\pi T_{\mu\nu}\ , \label{EE}
\ee
where $R_{\mu\nu}$ and $R$ are the Ricci tensor and scalar, and the right hand side is the effective  EMT. It includes the renormalized expectation value of all matter fields,  higher-order terms arising from its regularisation, and possible contributions arising from modifications to  Einstein-Hilbert gravity or a cosmological constant $\Lambda$.
However, our analysis does not use any specific property of the quantum state of matter and does not separate the matter EMT into the collapsing matter and (perturbatively obtained) quantum excitations \cite{MMT:22}.

	In discussing $\Phi$BH properties,  we apply the weakest form of  cosmic censorship and require the absence of scalar curvature singularities at the apparent horizon \cite{HE:73,FN:98}. In spherical symmetry it is enough to require that $G^\mu_{~\mu}\eqdef\mathrm{G}$ and $G_{\mu\nu}G^{\mu\nu}\eqdef\mathfrak{G}$ are finite \cite{MMT:22}. We express this observability by requiring a finite  formation time according to the  clock  of a distant Bob.

	A general spherically symmetric metric in Schwarzschild coordinates (with areal radius $r$) is given by
	\be
	ds^2=-e^{2h(t,r)}f(t,r)dt^2+f(t,r)^{-1}dr^2+r^2d\Omega_2\ , \label{sgenm}
	\ee
	while using the advanced null coordinate $v$ results in the form
	\be
	ds^2=-e^{2h_+(v,r)}f_+(v,r)dv^2+2e^{h_+(v,r)}dvdr+r^2d\Omega_2\ . \label{m:vr}
	\ee
	The function $f$ is coordinate-independent, i.e., $f(t,r)\equiv f_+\big(v(t,r),r\big)$, and in what follows we omit the subscript. It is conveniently represented via the Misner--Sharp--Hernandez (MSH) mass $M\equiv C/2$ \cite{F:15} as
	\be
	f=1-\frac{C(t,r)}{r}=1-\frac{C_+(v,r)}{r}\defeq\pad_\mu r \pad^\mu r\ ,
	\ee
	The functions $h$ and $h_+$ play the role of integrating factors in the coordinate transformation
	\be
	dt=e^{-h}(e^{h_+}dv-f^{-1}dr)\ . \label{trvr}
	\ee
	For example, the Schwarzschild metric corresponds to $h\equiv 0$, $C\equiv r_\sg=\mathrm{const}$, and $v=t+r_*$, where $r_*$ is the tortoise coordinate \cite{FN:98,HE:73}.

	In a cosmological setting   we assume that a separation of scales exists between geometric features associated with the black hole and those of the large-scale Universe \cite{DMSS:24,DSST:23}. In this case, the outer apparent horizons of the \pbh\ are given by the largest real root $r_\sg$ of $f(t,r)=0$, that is, belongs to  the near-region, i.e. $r_\sg\ll1/\sqrt{\Lambda}$.  Invariance of the MSH mass implies
	\be\label{msh}
	r_\sg(t)=r_+\big(v(t,r_\sg(t)\big)\ ,
	\ee
	where $r_+$ is the equivalent root of $f(v,r)$. Unlike the globally defined event horizon, the   apparent horizon is foliation-dependent. However, it is invariantly defined in all foliations that respect spherical symmetry \cite{FEFH:17} which is used in the following.

Both the analysis of the Einstein equations and the evaluation of curvature invariants are  conveniently performed using the   effective EMT components $\tau_a$, (where $a={}_t, {}^r, {}_t{}^r $)  are defined as \cite{MMT:22}
\begin{align}
	\tau{_t} \defeq e^{-2h} {T}_{tt}\ , \qquad {\tau}{^r} \defeq T^{rr}\ , \qquad  \tau {_t^r} \defeq e^{-h}  {T}{_t^r}\ . \label{eq:mtgEMTdecomp}
\end{align}
The Einstein equations for the components $G_{tt}$, ${G}{_t^r}$, and ${G}^{rr}$ are then, respectively,
\begin{align}
	\partial_r C &= 8 \pi r^2  {\tau}{_t} / f\ , \label{eq:Gtt} \\
	\partial_t C &= 8 \pi r^2 e^h  {\tau}{_t^r}\ , \label{eq:Gtr} \\
	\partial_r h &= 4 \pi r \left(  {\tau}{_t} +  {\tau}{^r} \right) / f^2\ . \label{eq:Grr}
\end{align}
To ensure finite values of the curvature scalars, it is sufficient to work with only two invariant quantities
\be
 {\mathrm{G} }\defeq\mathrm{T} +2T^\theta_{~\theta}, \qquad {\mathfrak{G}}\defeq\mathfrak{T}+2\big(T^\theta_{~\theta}\big)^2,
\ee
where
\begin{align}
	\mathrm{T} &= ( {\tau}{^r} -  {\tau}{_t}) / f\ , \\
	\mathfrak{T} & =\big( ( {\tau}{^r})^2 + ( {\tau}{_t})^2 - 2 ( {\tau}{_t^r})^2 \big) / f^2\ .
	\label{cond:2}
\end{align}
In our analysis, we can disregard the contributions of $T^\theta_{~\theta}\equiv T^\phi_{~\phi}$, as one can verify that they do not introduce further divergences \cite{MMT:22,MV:25}.

\subsection{Solutions}\label{1-sol}

There are two admissible classes of near-horizon solutions which are distinguished by the behaviour of the effective EMT components as $r\to r_\sg$, which scale as $f^k$ with $k=0,1$ \cite{MMT:22,MV:25}.
Solutions with $k=1$ describe eternal static configurations and the moments of formation and possible evaporation of a \pbh\ \cite{MMT:22,DSST:23}. Dynamics of a \pbh\ through its evolution is described by  a  $k=0$  solution, where  the three effective EMT components  scale  as
	\be
	\tau_t,\tau^r \to -\Upsilon^2, \qquad \tau_t^r\to\pm \Upsilon^2\ ,
	\ee
	 for some $\Upsilon(t)$.
	The two metric functions are
 \begin{align}
	C&= r_\sg-4\sqrt{\pi}r_\sg^{3/2}\Upsilon\sqrt{x}+c_1 x+\cO\big(x^{3/2}\big)\ , \label{Ctr}\\
h&=-\frac{1}{2} \ln{\frac{x}{\xi}}+h_{12}\sqrt{x}+\cO(x)\ , \label{htr}
	\end{align}
	where  $x\defeq r-r_\sg(t)$, and the function  $\xi(t)$ is determined by the  choice of time variable. As a result,
\be
f=\alpha_{12}\sqrt{x}+\cO(x), \qquad \alpha_{12}=4\sqrt{\pi r_\sg}\Upsilon .
\ee
 The higher order terms such as $c_1(t)$ and $h_{12}(t)$ are discussed in Appendix \ref{ap-trans:1}
	
	{As only two metric functions  --- $f(t,r)$ and $h(t,r)$ --- describe solutions of the Einstein equations, consistency of Eq.~\eqref{eq:Gtr} with Eqs.~\eqref{eq:Gtt} and \eqref{eq:Grr} requires that}
	\be
	\frac{d r_\sg}{dt} {\equiv r'_\sg} =\pm4\Upsilon\sqrt{\pi r_\sg\xi}=\pm\alpha_{12}\sqrt{\xi}\ ,      \label{lumin}
	\ee
	where the plus (minus) sign corresponds to the expansion (contraction) of the  outer  horizon.  The case of $r_\sg'<0$ is most conveniently described using the advanced null coordinate $v$. An evaluation of the expansions of null geodesic congruences \cite{HE:73,F:15} identifies the domain $f<0$ as a trapped region  and thus a \pbh. The case of $r_\sg'>0$ is most conveniently described using the retarded null coordinate $u$. In this case the domain $f<0$ is the anti-trapped region. For both the exterior boundaries are timelike hypersurfaces. Here we are concerned only with black holes.
	
	The function $-\Upsilon^2<0$ determines the energy density  at the outer apparent {(or anti-trapping)} horizon,  and  higher-order terms are matched with higher-order terms in the EMT expansion \cite{MMT:22}. Both solutions violate the null energy condition (NEC) {in the vicinity of the  horizon at $r=r_\sg$ }\cite{HE:73,KS:20}, i.e. there are future-directed null vectors $k^\mu$ such that $T_{\mu\nu}k^\mu k^\nu<0$.  This is consistent with the result that the apparent horizon is not ``visible" to a distant observer unless the NEC is violated \cite{HE:73,FN:98}.
	
In $(v,r)$ coordinates the black hole metric is described by
\begin{align}
	C_+(v,r) &= r_+(v)+w_1(v)y+\cO(y^2)\ , \label{c1}\\
	h_+(v,r) &= \zeta_0(v)+\zeta_1(v) y+\cO(y^2)\label{h1} \ ,
\end{align}
where $y\defeq r-r_+(v)$.
Note that a freedom in the redefinition of the null variable $v$ allows one to set $\zeta_0\equiv 0$. 		
From the definition of the apparent horizon it follows that $w_1\leqslant 1$. The inequality is saturated at the formation of a \pbh\  (more details can be found in \cite{DST:22}). We discuss relations between the expressions of  the metric functions in the two coordinate systems and behaviour of some of the expansion coefficients below.

The Schwarzschild sphere $r_\sg(t)$ is a timelike hypersurface \cite{MMT:22}. Therefore, ingoing null geodesics and some of the ingoing timelike geodesics  cross the apparent horizon in a finite time according to Bob. Similarly, the apparent horizon provides a natural redshift regulator in the membrane paradigm in Section~\ref{s:memb}.

The ingoing null geodesics satisfy
\begin{align}
 \left.\frac{dr}{dt}\right|_{v=\mathrm{const}}\!\!\!\!\!\!= -e^hf=r_\sg'-\ell_{12}\sqrt{x}- \frac{\ell_1}{r_\sg}x+\cO(x^{3/2}). \label{eHf}
\end{align}
Here the coefficient $\ell_{12}$ is expressed  via the redshift at the apparent horizon, $d\tau^2=\alpha^2dt^2|_\ah$, as
\be
\ell_{12}=\frac{\alpha^2}{2\sqrt{\xi}} ,
 \ee
and the coefficient $\ell_1$ is discussed below.
The  redshift  for a comoving observer at the apparent horizon, $r_\sg(t)\equiv r_\ah(\tau)$,   is given by
\begin{align}
\alpha^2&=\lim_{r\to r_\sg} \left(e^{2h}f^2-r_\sg'{}^2\right)/f \nonumber \\
&=\frac{2\xi}{r_\sg}(1-c_1+4\sqrt{\pi} h_{12}r_\sg^{3/2}\Upsilon) \label{redshift-def}
\end{align}

Using invariance of the MSH mass and Eq.~\eqref{trvr},  it is possible to obtain relations between  the coefficients of the metric function expansions in the two coordinate systems. For example,  the leading order relation between the coordinate distances from the horizon, $x\big(t(v,r),r\big)$ and $y$, is
\be
x=\half\omega^2 y^2, \qquad \ \omega^2=\frac{\alpha^4}{8\xi r_\sg'{}^2} . \label{x2y:d}
\ee
Appendix \ref{ap-trans:2} provides the details of the derivations of $x(y)$, $y(x)$ and related quantities.

There are several useful relations that hold on the apparent horizon. The coordinate-independent definition of the MSH mass and the timelike character of apparent horizon result in
\be
\left.\frac{dv}{dt}\right|_\ah=\frac{|r'_\sg|}{|r'_+|}=\frac{\alpha}{\sqrt{2|r'_+|}} .
\ee
 On the other hand, Eq.~\eqref{trvr} implies
\be
\left.\frac{dv}{dt}\right|_\ah=\frac{\ell_{12}}{\alpha_{12}}.
\ee

As we discuss in Section~\ref{s:kappa}  Hayward--Kodama surface gravity $\kappa_\mathrm{K}=(1-w_1)/(2r_+)$ is a natural generalisation of the surface gravity in stationary spacetimes. It coincides with the Schwarzschild value $\kappa=1/4M=1/2r_\sg$ only if $w_1=0$. % Moreover, assuming $w_1$ and taking the static limit as discussed in Section~\ref{sec-limit}, requires
%\be
%\xi=4\pi r_\sg^3\Upsilon^2, \label{xiUp}
%\ee
%without assuming the Page evaporation law,as it is done below (see Appendix \ref{ap-trans:3} for details).

We  use this value throughout, both because this is the basis for the Hawking temperature \cite{BD:84,BMPS:95}, and especially because  even small deviations of the first mass expansion coefficient from zero have dramatic influence on the quasinormal modes (QNM) spectrum \cite{MSST:25}.

Hence unless it is assumed otherwise we set $w_1=0$, which leads to
\be
\alpha^2=\frac{4\xi}{r_\sg}.
\ee
Assuming that $\kappa_\mathrm{K}=(2r_\sg)^{-1}$ and that the Page evaporation law \cite{FN:98,BMPS:95,P:05} has the same form in both $(t,r)$ and $(v,r)$ coordinates, with $r_\sg'=-A/r_\sg^2$, $r_+'=-A/r_+^2$, respectively, leads to the identification \cite{DSST:23}
\be
\Upsilon^2=\frac{A}{8\pi r_\sg^4}, \qquad \xi=\frac{A}{2r_\sg}. \label{Upxi}
\ee
It allows us to identify
\be
\alpha^2=\frac{4\xi}{r_\sg}=2|r_\sg'| , \qquad \ell_{12}=\frac{\alpha}{\sqrt{r_\sg}}. \label{al2rpr}
\ee
In addition, adapting the arguments of \cite{B:81} that in a steady state approximation of the black hole evaporation \cite{BMPS:95,FN:98} we obtain
\be
\zeta_1\sim |r_+'|/r_+ \ .
\ee
Hence, in the near horizon region $h_+\approx 0$, and  the  metric in $(v,r)$ coordinates as well approximated by a Vaidya metric with $2M(v)=r_+(v)$.

\subsection{Static limit}\label{sec-limit}
 Apart from a strikingly different form of the metric of Eqs.~ \eqref{Ctr} and \eqref{htr}, the differences from MBHs include violation of the NEC and the finite infall time according to a distant observer. Hence it is interesting and important to investigate potential observable consequences that follow from these differences. One technical difficulty is that Eqs~\eqref{eq:Gtr} and \eqref{cond:2} indicate that there are no static $k=0$ solutions. In fact, all static black hole models indeed belong to the class $k=1$.

However, outside the dramatic astrophysical events such as collisions \cite{LIGO:23,BCC:25,RZ:13}, ABHs are essentially stationary and modelled as such \cite{CP:19,RZ:13,FN:98,LISA:22}. For macroscopic objects, even as small as primordial black holes \cite{CS:18,Y:22} with high expected Hawking temperature, the dynamical evolution is slow \cite{A:23,IPS:23}. Hence we have to  put the \pbh\ metrics into the form that allows immediate comparison with their static counterparts.

We focus on the limiting form of the near horizon geometry, as this is the domain where the differences between the \pbh s and MBHs may be observed. From a mathematical point of view the limit  is naturally described in $(v,r)$ coordinates, where all metric parameters in Eqs.~\eqref{c1} and \eqref{h1} become constant, and $y\to x$. As majority of the static models have $h_+\equiv h=0$ we will assume this here. The transition to the static limit is conveniently analysed by assuming validity of the Page evaporation law and then taking the limit $A\to 0$.

The description of the transition in $(t,r)$ coordinates is more involved. The static metric function $f(r)$ belongs to $k=1$ class and has the near-horizon expansion \cite{MSST:25}
 \be
 f(r)=\sum_{k\geqslant 1} \frac{\bar\alpha_k}{r_\sg^k}x^k. \label{sta-f}
 \ee
 We are interested in a sufficiently smooth transition
 \be
  f(t,r)=\sum_{k\in\mathbb{Z}/2 } \frac{{\alpha}_k}{r_\sg^k}x^k \to f(r),
 \ee
 while $|r'_\sg|\to 0$.
That means in the expansion of $\sg^{rr}\equiv f=1-C(t,r)/r$ all   coefficients  of the half-integer powers $x^{(2j+1)/2}$ approach zero, and the coefficients of the integer powers approach the static values, e.g.,
\be
\alpha_1= 1-c_1\to\bar\alpha_1, \qquad  \alpha_2=-1+c_1-c_2r_\sg\to \bar\alpha_2.
\ee
Indeed, $r'_\sg\to 0$ and additional relations of the previous Section imply that $\Upsilon\to0, \xi\to 0$ individually, while the appropriate behaviour of the higher order EMT components ensures a smooth approach to the static limit.

On the other hand, using   Eq.~\eqref{htr} to write
\be
e^h=\sqrt{\frac{\xi}{x}}\exp\Big(\sum_{j\geqslant1}h_{j/2}x^{j/2}\Big)\eqdef \sqrt{\frac{\xi}{x}}e^{\bar h} , \label{hbarh}
\ee
clearly shows that the static limit of the series expansion that necessarily involves $\xi\to 0$ is singular, with the coefficients in the expansion of $\bar h$  diverging as powers of $1/\xi$. Hence
we demand only that in the asymptotically flat case $e^h\to1$, so
\be
e^{h(t,r)}f(t,r)\to f(r), \qquad e^{2h(t,r)}f(t,r)\to f(r),
\ee
and use the approximations that satisfy this limit and give the correct form of a sufficient number of the expansion coefficients $h_{j/2}$ at finite $r'_\sg$. We discuss the approximate form of $e^h$ in Appendix \ref{ap-trans:3}

The coordinate transformation of Eq.~\eqref{trvr} becomes the standard transformation for the Eddington--Finkelstein coordinates, $dv=dt+dr/f$. The first two terms in the expansion $e^h f$ of Eq.~\eqref{eHf} go to zero as $r'_\sg\to 0$ (again, conveniently modelled by $A\to 0$), and $e^h\to 1$ means $\ell_1\to\alpha_1$.

% We also show how dominant near horizon contribution to the finite radial propagation time between the two points   according  to the distant observer, $\delta t\sim (x_2-x_1)/|r_\sg'|$ transitions to the logarithmic %divergence $\Delta t\sim r_\sg \ln (x_1/x_2)$ with $x_2\to 0$.

It is also straightforward to see that only the leading half-integer order terms are of consequence for any   $A<1$. For nonzero $r'_\sg$ it is also possible to study motion of the test particles or response to gravitational perturbations by ``freezing" the metric, as the characteristic timescales of these processes are much shorter than the \pbh\ characteristic time $r_\sg/|r'_\sg|\sim r_\sg^3/A$. Indeed, taking $\alpha_1\sim 1$ we get the estimate of the range of the coordinate distance from the apparent horizon where the deviations of the frozen metric that approximates the \pbh\, geometry via Eq.~\eqref{fro-f} from the static metric of Eq.~\eqref{sta-f} by identifying $x_*$ for which the first two terms of Eq.~\eqref{fro-f} coincide,
\be
\frac{x_*}{r_\sg}=\alpha^2\sim|r_\sg'|.
\ee
For the same value $x=x_*$ the contribution of the next term $\alpha_{3/2} x^{3/2}/r_\sg^{3/2}$ is smaller by the factor of $\alpha$.

Travel time up $x_*$ is well approximated by the standard classical result. Eq.~\eqref{eHf} then indicates that the time that it takes to complete the infall is $\Delta t_*=\cO(r_\sg)$. Thus the natural \pbh\, cutoff on a blueshift $1/\alpha$ ensures that the actual infal takes of the order of the black-hole light-crossing time $r_\sg$.

As a result, the simplest explicit forms of the \pbh-modified metric functions that satisfy all these requirements are given by
 \begin{align}
	&f=\alpha_{12}\sqrt{x}+\frac{\alpha_1}{r_\sg} x+\cO(x^2), \label{fro-f}\\
&e^h=\sqrt{b^2+\frac{\xi}{x}}+d \label{fro-h} ,
	\end{align}
where $\alpha_{12}= 4\sqrt{\pi r_\sg}\Upsilon$, $b+d=1$ (or another asymptotic value),  and $h_{12}=d/\sqrt{\xi}$.

\section{Apparent horizons as membranes}\label{s:memb}
\pbh s provide a natural regulator for the redshift and the freefall acceleration in the frame of the observer at the horizon. This suggests treating their apparent horizons as membranes \cite{TPM:86}, and below we derive some of their basic properties. Relationships between the near-horizon geometry of a black hole and the Rindler geometry are profound, intriguing, and potentially important. They can be put on a rigorous basis for \pbh s if a particular hypersurface, known as the York--Frolov separatrix, is used as a baseline \cite{DS:23}. In Section~\ref{s:memb2} we show that it is also a hypersurface of approximately zero redshift.

\subsection{Basic membrane properties}\label{s:memb1}

In classical general relativity, freely falling observers experience no special event as they cross the event horizon of an MBH. However, to a distant static observer (Bob), an infalling observer (Alice) appears to be frozen at the horizon due to the infinite redshift. Hence, the BH interior can be regarded as irrelevant for a distant observer. The membrane paradigm \cite{TPM:86} is based on this complementary picture. Accordingly, a distant observer excludes the interior of an MBH by a fictitious timelike membrane (also known as a stretched horizon), located at some coordinate distance $\epsilon$ outside the horizon. Physical quantities accessible to Bob are independent of $\epsilon$, at least at leading order. Identifying the membrane with the timelike apparent horizon removes this ambiguity altogether.

The two most basic quantities that are needed to characterise this surface at some $r=r_\sg+\epsilon$, $\epsilon\ll r_\sg$ \cite{TPM:86} are the redshift and the freefall acceleration that is experienced by a static observer (Eve) there.  For a hairless static MBH of Section ~\ref{sec-limit}, these are
\be
\alpha=\sqrt{f(r)}, \qquad g=\frac{f'(r)}{2\sqrt{f}}\to\frac{\kappa}{\alpha}, \label{al-sta}
\ee
respectively, where $\kappa$ is the surface gravity. We discuss different dynamical  generalisations of the surface gravity in Section~\ref{s:kappa}.

Using the Israel junction conditions \cite{P:04} the membrane is endowed with the EMT of a 2D  viscous fluid whose physical properties are such that it has the same phenomenology as the BH it represents \cite{TPM:86,AAOW:20,MBMP:20,CMMP:22}. A careful analysis of membrane's properties led to Ohm’s law, Joule’s law, and the non-relativistic Navier–Stokes equation.

In a fully classical picture there is a peculiar relations between the shear viscosity $\eta$ and the bulk viscosity $\zeta$,
\be
\eta_\mbh=-\zeta_\mbh=\frac{1}{16\pi}.
\ee
In general, viscosity depends on quantum corrections \cite{AAOW:20,CMMP:22,SM:24,SMCP:25}

The Einstein equations governing the metric perturbations can be reduced to two Schrödinger-like master equations: the Regge–Wheeler equation (describing axial perturbations) and the Zerilli equation (describing polar perturbations) \cite{FN:98}.
Using the membrane as a proxy for the black hole boundary, imposing boundary conditions on it and at infinity defines the eigenvalue problem for these equations. The boundary conditions  involve both viscosities for the polar perturbations and only the shear viscosity for the axial ones \cite{SM:24, SMCP:25}. In this case the membrane reflectivity \cite{AAOW:20,MBMP:20,CMMP:22} is given by
\be
|\mathcal{R}|=\left|\frac {1-\eta/\eta_\mbh}{1+\eta/\eta_\mbh}\right|.
\ee
%and is thus directly tied to the boundary conditions that are imposed in black hole perturbation problems.

The \pbh\ framework provides a natural candidate for the membrane -- the contracting  timelike apparent horizon. As the $(v,r)$ coordinates are regular across the horizon we mostly use them for the analysis below, but  physically relevant quantities can be extracted also using the Schwarzschild coordinates.

The results of Section \ref{1-sol} indicate that the redshift $\alpha\sim\sqrt{2|r_\sg'|}$.  Using $d\tau^2=\alpha^2_vdv^2|_\ah$  we find
\be
\alpha_v^2=2|r'_+(v)|.
\ee
The apparent horizon is a timelike hypersurface given by an implicit equation $\Phi=r-r_+(v)=0$. Its spacelike unit normal $n_\mu\propto \pad_\mu \Phi$. The four-acceleration $a^\mu=Du^\mu/d\tau$ is proportional to it,
\be
a^\mu=g_v n^\mu, \qquad n_\mu=-\frac{1}{\sqrt{2}}\left(\sqrt{|r_+'|},1/\sqrt{|r_+'|},0,0\right).
\ee
Its magnitude is
\be
g_v=-\frac{(1-w_1)r'_+ +2\zeta_1 r_+r'^2_++r_+r''_+}{2\sqrt{2}r_+|r'_+|^{3/2}},
\ee
which on the approach to the static limit (Section \ref{sec-limit}) becomes
\be
g_v\to \left( 2\sqrt{2|r'_+|}r_+)\right)^{-1}. \label{gH}
\ee

The apparent horizon is coordinatised by the proper time $\tau$ and the angular variables $\theta$ and $\phi$. The orthogonal triad on the apparent horizon $e_{a}^\mu$, $a=0,2,3$ is formed by the four-velocity $u^\mu$ of a comoving observer and the two tangent vectors $\pad_\theta$ and $\pad_\phi$, respectively. We comment in passing that in the original spirit of the membrane paradigm the acceleration of a freely falling Alice in the reference frame of the comoving Eve on the apparent horizon is $-g_v$, directed ``downwards".

The extrinsic curvature  $K_{ab}=n_{(\mu;\nu)}e_{a}^\mu e_{b}^\nu$ has  diagonal form,
\be
K^{a}_{~b}=\mathrm{diagonal}\left(g_v, \frac{\alpha_v}{2r_+},\frac{\alpha_v}{2r_+}\right),
\ee
where the dependence on $v=v(\tau)$ is assumed. We contrast it with the extrinsic curvature tensor in a static spherically symmetric gravitational field with $h\equiv0$,
\be
K^{a}_{~b}=\mathrm{diagonal}\left(g, \frac{\alpha}{r},\frac{\alpha}{r}\right),
\ee
which is evaluated for a surface at $r=\mathrm{const}$.

A fictitious matter distribution on the membrane is obtained by formally postulating discontinuity in transversal metric components between the two bulk regions, outside ($\cal{M}^+$) and inside ($\cal{M}^-$) of the membrane. Following \cite{TPM:86} we set $K^+_{ab}:=K_{ab}$, $K^-_{ab}\equiv 0$ (an alternative convention \cite{AAOW:20,SMCP:25} is $K^-_{ab}=-K^+_{ab}$). The hypersurface EMT
 $S^{a}_{~b}$ is introduced via the Israel junction condition  \cite{P:04},
\be
 [K]\delta^a_b -[K^{a}_{~b}]=8\pi T^{a}_{~b},
\ee
where $[X]$ indicates the difference between the outside and the inside values of the   quantity $X$ on its approach to the membrane. In a general case it has the form corresponding to a two-dimensional dissipative fluid \cite{TPM:86,AAOW:20},
\be
T_{ab}=\rho u_au_b+(p-\zeta\vartheta)\gamma_{ab}-2\eta\sigma_{ab}.
\ee
The projector tensor $\gamma_{ab}\defeq h_{ab}+u_a u_b$ is defined using the induced metric $h_{ab}\defeq \sg_{\mu\nu}e^\mu_a e^\nu_b$ on the membrane. For the $h_{ab}$  metric compatible covariant derivative $D_a$ the  expansion and the shear are defined as $\vartheta=D_a u^a$  and $\sigma_{ab}\defeq\half( \gamma^a_bD_c ub+\gamma^c_d D_cu_b-\vartheta \gamma_{ab})$. In contrast to the static background (where both the shear and the expansion are zero), here
\be
\vartheta=\frac{2\dot r_+(\tau)}{r_+(\tau)}=-\frac{\alpha_v}{r_+}, \qquad \sigma_{ab}=-\frac{\vartheta}{2}\gamma_{ab}.
\ee
If one still accepts $\eta=-\zeta=1/(16\pi)$, then the two-dimensional density and pressure are
\begin{align}
&\rho=-\frac{\alpha_v}{8\pi r_+}, \\
& p=\frac{1}{8\pi}\left(g_v+\frac{3}{2}\frac{\alpha_v}{r_+}\right)\approx\frac{1}{6\pi r_+}\left(\frac{1}{\alpha_v}+3\alpha_v\right).
\end{align}
Analogous to the standard membranes, in the static limit $r_g'\to0$ energy surface density goes to zero while the two-dimensional pressure diverges with $g_v$.

We can formally define the speed of sound on the membrane $c_s=\sqrt{\pad p/\pad \rho}$, \cite{MNT:18,MBMP:20}. To express the isoentropic condition \cite{RZ:13} we treat both quantities as functions of $\alpha_v$ (i. e. the evaporation rate), and keep the Schwarzschild radius constant. Then
\be
c_s\approx \frac{1}{2\alpha_v^2}\gg 1,
\ee
diverging in the static MBH limit, also similarly to the standard membrane case.

\subsection{Separatrix}\label{s:memb2}

An elegant mathematical relation between the near-horizon Schwarzschild (or Kerr) metric and the Rindler metric \cite{BD:84,BMPS:95} is useful for the anlysis of the Hawking radiation and serves as a starting point of many aspects of gravitational thermodynamics \cite{BMPS:95,J:95,P:02,P:10}. In the Schwarzschild case near the event horizon the two-dimensional part of metric is
\begin{align}
	ds^2\approx -\frac{x}{r_\sg}dt^2+\frac{r_\sg}{x}dx^2=-\kappa^2\mathfrak{x}^2dt^2+d\mathfrak{x}^2+dL_\perp^2, \label{SR-0}
\end{align}
where a new independent variable $\mathfrak{x}$ corresponds to the physical distance from the horizon,
\be
\mathfrak{x}(x)=\int_{r_\sg}^{r_\sg+x}\frac{dr}{\sqrt{f(r)}},
\ee
and $\kappa$ is the surface gravity. The association of $\kappa$ with the acceleration and $(t,\mathfrak{x})$ with the Rindler coordinates allows the standard conformal mapping into the two-dimensional Minkowski spacetime.

Despite importance of the Rindler--Schwarzschild relation, its rigorous establishment is non entirely straightforward. Expansion near the apparent horizon of a \pbh\ does not lead to anything resembling the right hand side of Eq.~\eqref{SR-0}. The key insight of Ref.~\cite{DS:23} was to take a special hypersurface --- the York--Frolov separatrix \cite{Y:83,F:16,BHL:18} as the base.

It was originally introduced  \cite{Y:83} to provide an approximate locally derived expression for the event horizon. The event horizon is the boundary between outgoing null geodesics that reach future null infinity and those that do not; if it forms, in an ordered family $R(v)$, it is generated by the first family that fails to escape. For a future-directed outgoing family of null geodesics, one of which family we denote $R(v)$, at the apparent horizon of a \pbh\
 both expansion $\vartheta_{{\mathrm{out}}}=0$ and $R'(v)=0$ are zero. The photons are only momentarily at rest and escape to finite distances  in finite (advanced) time. On the other hand, the event horizon generators are photons that are ``stuck'', which can be quantified as $d^2R/dv^2=0$. Thus the solution $r_\mathrm{sep}(v)$ of the algebraic equation
\begin{align}
	2\frac{d}{dv}(e^hf)+e^h f\frac{d}{dr}(e^hf)=0 , \label{YFsep}
\end{align}
provides a good approximation for the location of the event horizon.

For low luminosity $L \ll 1$, a perturbative analysis establishes that $r_\mathrm{sep}(v)$ is close to $r_+$. Thus we obtain the leading-order expression for $y_\mathrm{sep} \defeq r_\mathrm{sep}(v)-r_+(v)$ for a sufficiently small $w_1$,
\begin{align}
	y_\mathrm{sep}\cong 2r_+(1+w_1)r'_+. \label{app:sep}
\end{align}
while a more general expression is given in Appendix \ref{a-sep}.

Event horizons are absent in regular black holes. In many of their models it is still a useful distinction between null and timelike geodesics that can leave the trapped region only at its final disappearance  and those that can cross the timelike apparent horizon before the evaporation is complete. The boundary of the two domains is a null geodesic (named a D-geodesic \cite{BHL:18}). The condition of  Eq.~\eqref{YFsep} is much easier to implement, and the resulting   separatrix  $r_\mathrm{sep}(v)$  is a good approximation to it.

The separatrix plays a role similar to the event horizon also in its another aspect. In a static ($k=1$) case the redshift of a static Eve approaches zero according to Eq.~\eqref{al-sta} when her position approaches the event horizon. For slow evolving \pbh\ with $\zeta\sim|r'_+|/r$ and $w_1\ll 1$ for an observer at $r_\mathrm{sep}(v)$ that moves with the same (coordinate) velocity as the  comoving horizon observer, $dr/dv=r_+$, the redshift is approximately  zero. Indeed, using Eq.~\eqref{m:vr} we find
\be
d\tau^2=\cO(w_1^2|r'_+|, |r'_+|^2) dv^2 .
\ee

\section{Surface gravity}\label{s:kappa}

The surface gravity $\kappa$ plays an important role  particularly in black hole thermodynamics and more generally in semiclassical gravity \cite{HE:73,BD:84,BMPS:95,AK:04}. For an observer at infinity the Hawking radiation that is produced on the background of a stationary black hole is thermal with its temperature given by $\kappa/2\pi$. However, surface gravity is unambiguously defined only in stationary spacetimes, where there are several equivalent definitions.

Stationary asymptotically flat spacetimes  admit a Killing vector field $\xi^\mu$, $\xi^\mu\xi_{(\mu;\nu)}=0$,  that is timelike at infinity \cite{HE:73,F:15}. A Killing horizon is a hypersurface on which the norm $\sqrt{\xi^\mu\xi_\mu}=0$. While logically this concept is independent of the notion of an event horizon, the two are related: in a stationary asymptotically flat spacetime the event horizon coincides with the Killing horizon \cite{FN:98}. The Killing property $\xi_{(\mu;\nu)}=0$ results in $\xi^\mu\xi_\mu=\mathrm{const}$ on each of its integral curves, and the surface gravity $\kappa$ can be introduced as the inaffinity of null Killing geodesics on the event horizon,
\begin{align}
	\xi^\mu_{~;\nu}\xi^\nu \defeq \kappa \xi^\mu.
\end{align}

On the other hand, assuming sufficient regularity of the metric, expansion of the null geodesics near the apparent horizon $r > r_\sg$   establishes the concept of peeling affine gravity \cite{VAC:11,CLV:13},
\begin{align}
	\frac{dr}{dt} = \pm 2 \kappa_\mathrm{peel}(t) x + \mathcal{O}(x^2). \label{peeld}
\end{align}

Finally, $\kappa$ can be intuitively described as the force that would be required by an observer at infinity to hold a particle (of unit mass) stationary at the event horizon.  For a static observer Eve at some fixed areal radius $r$ the square of her four-acceleration $g^2_\mathrm{E}$  satisfies
\begin{align}
	g_\mathrm{E}(r) \defeq \sqrt{a^\mu_{\mathrm{E}} a_{{\mathrm{E}}\mu}} = \frac{r_\sg}{2r^2\alpha_\mathrm{E}},
\end{align}
where the redshift $\alpha_\mathrm{E}\equiv\sqrt{|\sg_{00}|}=e^h\sqrt{f}=\sqrt{1-r_\sg/r}$.  Correcting for it and taking Eve's position to the horizon  results in the surface gravity,
\begin{align}
	\kappa=\lim_{r\to r_\sg}\alpha_\mathrm{E} g_\mathrm{E}=\half f'(r_\sg)=1/(2r_\sg). \label{kappa-force}
\end{align}
(for a general static metric the redefinition of time allows to set $h(r_\sg)=0$, and the last equality is valid for the Schwarzschild metric).

Generalisations of these definitions to arbitrary spacetimes are inequivalent, but for slowly evolving macroscopic black holes are thought to give close results \cite{VAC:11,NV:06}. In spherically-symmetric spacetimes Kodama vector field \cite{K:80} has many properties of the Killing field and results in the expression for surface gravity \cite{H:98,F:15}
\be
\kappa_\mathrm{K}=\frac{1}{2}\left.\left(\frac{C_+}{r^2}-\frac{\pad_r C_+}{r}\right)\right|_{r=r_+}=\frac{(1-w_1)}{2r_+}.
\ee
It is well-defined for \pbh s and if one requires that its value corresponds to the standard surface gravity (which is necessary for the validity of the black hole evaporation seen as a sequence of the Schwarzschild background snapshots  \cite{BMPS:95,FN:98}), then $w_1=0$ must hold.

  A naive application of Eq.~\eqref{kappa-force} leads to a divergent quantity \cite{MMT:22}, seemingly invalidating the relationship between the surface gravity and acceleration at the horizon. Below we show how it can be correctly extended to dynamical spacetimes.

There are several versions of the peeling surface gravity that run into difficulty because their definitions presuppose at least continuity of the function $h(t,r)$ on the apparent horizon.  The only definition that does not trivially result
in zero or infinity \cite{MMT:22s} was introduced using flat slice Painlev\'{e}--Gullstrand coordinates $(\bt,r)$ \cite{NV:06} (whose relevant properties are  summarized in Appendix \ref{ap-PG}),
\be
	\kappa_{\PG_2}\defeq\left.\frac{1}{2r_\sg}(1-\pad_r \bar{ C}+\pad_{\bt} \bar{C})\right|_{r=r_\sg}, \label{kpg2}
\ee
where $\bar C(\bar t,r)=2M\big(t(\bar t, r),r \big)$ is the MSH mass expressed in the Painlev\'{e}--Gullstrand (PG) coordinates. It was shown \cite{MMT:22s} that
\be
	\kappa_{\PG_2}=\left.\frac{ \pad_\bt \bar{C}}{2r_\sg}\right|_{r=r_\sg},
\ee
where
\be
	\pad_\bt\bar{C}=\pad_t C\pad_\bt t|_r.
\ee
Thus
\begin{align}
	\pad_\bt \bar{C}\approx\frac{r'_\sg}{\pad_t\bt}\left(1+\frac{2\sqrt{\pi r_\sg^3}\Upsilon}{\sqrt{r-r_\sg}}\right),\label{mess1}
\end{align}
and the behavior of the function $\bt(t,r)$ near the apparent horizon determines the limit. As we found  $\pad_t\bt\approx 1$ in its vicinity (see Appendix \ref{ap-PG}), this version of the surface gravity is also untenable for a \pbh.

On the other hand, if we modify  the intuitive description of $\kappa$ by replacing the static observer Eve with the one comoving with the contracting apparent horizon, for a slowly evolving \pbh\
\be
 \lim_{r\to r_+}\alpha_v g_v\approx \frac{1-w_1}{2r_+} =\kappa_\mathrm{K},
\ee
where $g_v$ is given by Eq.~\eqref{gH}, and the Schwarzschild limit corresponds to $w_1=0$.

\section{Discussion}\label{s:dis}

Despite the main $k=0$ family of solutions not having static metrics, it is possible to study their static limit in which the $k=1$ MBHs are approached. The resulting frozen $k=0$ metric of Eqs.~\eqref{fro-f} and \eqref{fro-h} is a suitable starting point for the analysis of the quasinormal modes (QNMs) and light rings of \pbh s. The near- and far-region geometries can be connected using Pad\'{e} approximants \cite{RZ:14,KRZ:16,KZ:22}. The $\alpha_{12}\sqrt{x}$ term easily adds on the interpolation scheme of Ref.~\cite{MSST:25} as  $\alpha_{12}\sqrt{x}/(1+x^p)$, for a sufficiently large integer $p$, while the approximation to $e^h$ is used directly.

Independently of the slow-evolution assumption, the first steps in implementing the membrane paradigm and a consistent description of the boundary conditions in terms of a 2D fluid on a timelike apparent horizon are carried out: local redshift, proper acceleration, relevant extrinsic data, and a viscous-fluid EMT are obtained. These data parameterise dissipation/reflectivity and can be used to assess their impact on quasinormal spectra and possible echoes at any rate of dynamics.

For surface gravity, the remaining peeling-type generalisation is ruled out for \pbh s. On the other hand, the intuitive interpretation of the surface gravity in stationary spacetimes ---- proper acceleration of an observer just outside the horizon, redshifted to infinity --- is naturally extended to dynamical \pbh s. It is close to the well-defined  Kodama--Hayward surface gravity. The latter coincides with the momentarily Schwarzschild value $\kappa=1/(4M)$ only when the expansion parameter $w_1=0$.

Our results provide  the necessary background for the quantitiative evaluation of the light rings, QNMs and possible echo structures for \pbh s and their comparison with the standard results in spherical symmetry. This will be the subject of the subsequent works.

.

		\acknowledgements
		 This work was supported by the ARC Discovery project Grant No. DP210101279 and by the Schwinger Foundation. I am grateful to the Perimeter Institute for hospitality, Niayesh Afshordi, Viqar Husain, Jose Lemos  and Carlo Rovelli for stimulating discussions, and Swayamsiddha Maharana and Rama Vadapalli  for helpful comments.
		
\appendix

\section{EQUATIONS AND SOLUTIONS}\label{ap-trans}

\subsection{EMT structure and the Einstein equations} \label{ap-trans:1}
The effective EMT components of $k=0$ \pbh s outside the Schwarzschild radius $r_\sg$ has the form
		\begin{align}
		&\tau_{t}=-\Upsilon^2+e_{12}(t)\sqrt{x}+e_{1}(t)x+\cO(x^{3/2})\ ,\label{tau-t}\\
		&\tau^{r}_{t}=-\Upsilon^2+\phi_{12}(t)\sqrt{x}+\phi_{1}(t)x+\cO(x^{3/2})\ ,\label{tau-tr}\\
		&\tau^{r}=-\Upsilon^2+p_{12}(t)\sqrt{x}+p_{1}(t)x+\cO(x^{3/2})\ ,\label{tau-r}
		\end{align}
where
\begin{align}
		\phi_{12}=\frac{1}{2}(e_{12}+p_{12})\ .
		\end{align}
The two subleading  coefficients are
\begin{align}
&c_1=\left(\frac{1}{3}+\frac{4 \sqrt{\pi} e_{12}r_\sg^{3/2}}{3\Upsilon}\right) , \label{c1} \\
&h_{12}= \left(\frac{1}{3\sqrt{\pi}r_\sg^{3/2}\Upsilon}-\frac{e_{12} - 3 p_{12}}{6\Upsilon^2}\right). \label{h12}
\end{align}
We obtain higher  order expressions by comparing expansions of the Einstein equations \eqref{eq:Gtt}--\eqref{eq:Grr}. We also quote the coefficient $\ell_1$ of Eq.~\eqref{eHf},
\begin{align}
\ell_1=&- \sqrt{\xi} \big[c_{32}r_\sg-(1-c_1)h_{12}r_\sg \nonumber \\
&+2\sqrt{\pi r_\sg}\,\Upsilon(2-(2h_1+h_{12}^2)r_\sg)\big]
\end{align}

In the $(v,r)$ coordinates convenient  EMT components are obtained from $\Theta_{\mu\nu}$ as
		\begin{align}
		\theta_{v}=e^{-2h_{+}}\Theta_{vv}\ ,\quad \theta_{vr}=e^{-h_{+}}\Theta_{vr}\ ,\quad \theta_{r}=\Theta_{rr}\ .
		\end{align}
Using  the coordinate transformation Eq.~\eqref{trvr} one can find relations between the effective EMT components in $(v,r)$ with those in $(t,r)$,
		\begin{align}
		\theta_{v}=\tau_{t},\quad \theta_{vr}=\dfrac{\tau^{r}_{t}-\tau_{t}}{f}\ ,\quad \theta_{r}=\dfrac{\tau_{t}+\tau^{r}-2\tau^{r}_{t}}{f^{2}}\ .\label{eff-EMT-tr-vr}
		\end{align}

The Einstein equations then take the following form
		\begin{align}
		&\partial_{v}C_{+}=8\pi r^2(\theta_{v}+f\theta_{vr})\equiv 8\pi r^2\Theta^v_{~v} ,\label{dvCp}   \\
		&\partial_{r}C_{+}=-8\pi r^2\theta_{vr} \equiv -8\pi r^2 \Theta^r_{~v}\label{drCp}\ ,\\
		&\partial_{r}h_{+}=4\pi r\theta_{r}\equiv 4\pi r e^{h_+}\Theta^v_{~r}\ .
		\end{align}
Hence for solutions with $h_+\equiv 0$  the relation $\tau_{t}+\tau^{r}=2\tau^{r}_{t}$ holds identically.	
	
\subsection{Expansion coefficients and gap functions}\label{ap-trans:2}		
		
	Expanding the lhs of Eq. \eqref{drCp} in a series around $r_{+}$ and the rhs around $r_{g}$, after making use of Eq. \eqref{eff-EMT-tr-vr}, and comparing order-by-order, one arrives at the following relation for $w_{1}(v)$:
		\begin{align}
		w_{1}(v)=\frac{e_{12}-p_{12}}{\Upsilon}\sqrt{\pi}r^{3/2}_{g}\label{w1-t}
		\end{align}
		The condition $e_{12}(t)=p_{12}(t)$ is therefore equivalent to $w_{1}(v)=0$.
Using Eqs.~\eqref{redshift-def},\eqref{c1} and \eqref{h12} we can rewrite the redshift as
\be
\alpha^2=\frac{4\xi}{r_\sg}(1-w_1) .
\ee

 Explicit leading order relations between different coordinates in the vicinity of the apparent horizon are based on Eq.~\eqref{trvr}. Starting from a point on the apparent horizon, $r_+(v)=r_\sg\big(t(v,r_+(v)\big)$, we relate the leading order coordinate differences $x=r-r_\sg(t)$ and $y=r-r_+(v)$.

 First, consider the constant advanced coordinate. 	
		We thus evaluate the change in  $t$ from the value $t\big(v,r_+(v)\big)$ along an ingoing null geodesic $v=\mathrm{const}$.  Along such a geodesic the time $t(v,r)$ varies as
		\be
			t(v,r_++y)=t(v,r_{+})+\left.\partial_{r}t\right|_{r_+}y+\tfrac{1}{2}\left.\partial^2_{r}t\right|_{r_+}y^2+\cO(y^3).
		 \ee
		Determining the explicit form of the above relation requires evaluating partial derivatives at the apparent horizon. Eqs.~\eqref{trvr} or \eqref{eHf} implies
		\begin{align}
			\partial_{r}t=-e^{-h(t,r)}f(t,r)^{-1}=\frac{1}{r'_{g}}+\cO(\sqrt{x})\ .
		\end{align}
		 The time variation  $\delta t\defeq	t(v,r_{+}+y)-t(v,r_{+})\,$ along an ingoing null geodesic is thus given by
		\begin{align}
			\delta t=\frac{y}{r'_{g}}+\tfrac{1}{2}(\partial^2_{r}t)\!\!\!\underset{y=0}{\big|}y^2+\cO(y^3)\label{delta-t}.
		\end{align}

The corresponding expansion of the  Schwarzschild radius $r_{\sg}(t)$  is given by
		\be\begin{aligned}
			r_{\sg}(t(v,r_{+}+y))&=r_{\sg}(t(v,r_{+}))+r'_{\sg}\delta t\\
			&\quad+\tfrac{1}{2}r''_{\sg}\delta t^2+\cO(\delta t^3)\ , \label{rg-var}
		\end{aligned}\ee
	  where keeping terms of order $\delta t^2$ is crucial.

		The coordinate distance $x(t,r)=r-r_\sg(t)$ is expressed as a function of the advanced null coordinate $v$ and $r$,
		\be
 x(v,r_{+}+y)=(r_{+}+y)-r_{g}(t(v,r_{+}+y)) . \label{x-gen}
		\ee
Using Eqs.~\eqref{delta-t} and \eqref{rg-var} in \eqref{x-gen} along with the invariance of the MSH mass \eqref{msh} then results in the quadratic relationship between $x$ and $y$ near the apparent horizon:
	 \begin{align}\label{xy}
	 	x=\tfrac{1}{2}\, \omega^2y^2 \ ,\quad\text{where}\quad  \omega^2\equiv-r'_{g}(\partial^2_{r}t)\!\!\!\underset{y=0}{\big|}\!\!-\frac{r''_{g}}{(r'_{g})^2}
	 \end{align}

The second derivative  of $t$ over $v$ is
\begin{align}
\pad^2_v t(v,r)& =\pad_v t\pad_t e^{-h+h_+} \nonumber \\
&= \sqrt{\frac{x}{\xi}}e^{-2(\bar h-h_+)}\left( -\frac{r_\sg'}{2\sqrt{\xi x}} +\cO\big(\sqrt{x}\big)\right), % \nonumber \\
%&\to -\frac{r_\sg'}{\xi\sqrt{2}}
\end{align}
that at the apparent horizon becomes
\be
\pad^2_v t(v,r_+)=\frac{|r'_\sg|}{2\xi}\to \frac{1}{r_\sg},
\ee
where the last expression on the rhs is valid when $w_1=0$ due to Eq.~\eqref{al2rpr}.

The second derivative  of $t$ over $r$ is
\be
\pad_r^2t(v,r)=\frac{1}{(e^hf)^2}\left(\pad_r t\pad_t+\pad_r\right)e^hf
\ee
and using Eq.~\eqref{eHf} we find
\be
\pad_r^2t(v,r_+)=-\frac{r_\sg''}{(r_\sg')^2}-\frac{\ell_{12}^2}{2 (r_g')^3}.
\ee
This enables to give the explicit form to the parameter $\omega$,
as
\be
\omega^2=\frac{\ell_{12}^2}{2(r_\sg')^2}\to\frac{1}{2\xi} ,
\ee
when the last expression is again valid for $w_1=0$.

 Then, using invariance of the MSH mass we observe
\begin{align}
	C_+(v,r) &= r_+(v) + w_1 y + \ldots = C \big( t(v,r),r \big) \nonumber \\
	&= r_\sg \big( t(v,r_+) \big) + r_\sg' \left( \frac{y}{r_\sg'} \right) -  {2 \sqrt{2 \pi r_\sg^3} \Upsilon \omega} y + \ldots \nonumber \\
	&=r_++\left( 1 -  {2 \sqrt{2 \pi r_\sg^3} \Upsilon \omega} \right) y+\ldots,
\end{align}
and find
\begin{align}
	w_1 = 1 - 2 \sqrt{2 \pi r_\sg^3} \Upsilon \omega .% \; \substack{{\scriptscriptstyle k=0}\\=} \; \frac{\sqrt{\pi}}{\Upsilon} r_\sg^{3/2} \left( e_{12} - p_{12} \right) , \label{w1t}
\end{align}

Now we obtain the leading term of the inverse transformation. For the constant time $t$ integration of Eq.~\eqref{trvr} leads to the leading order expression
\be
 \delta v(x)=\frac{\sqrt{x}}{2\sqrt{\pi r_\sg} \Upsilon }.
\ee
Following the radius outwards at  $t=\mathrm{const}$   results  in the relation $y(x)$,
\begin{align}
y(t,r_\sg+x)=r_\sg+x-r_+(v+\delta v)=\frac{|r'_+|\sqrt{x}}{2\sqrt{\pi r_\sg} \Upsilon } +\cO(x) .
\end{align}
Expanding $C_+\big(v(t,r),r\big)=C(t,r)$ results in the identities such as $r_+'(1-w_1)=-8\pi r_+^2\Upsilon^2$ that follow from the transformation law of Eq.~\eqref{eff-EMT-tr-vr} and the Einstein equation \eqref{dvCp}.

%{where the rightmost expression that is valid for $k=0$ solutions is obtained using Eqs.~\eqref{eq:vrEEvr} and the limiting form of Eq.~\eqref{thevr} close to the horizon, and $e_{12}$ and $p_{12}$ denote the %$\mathcal{O}(\sqrt{x})$ coefficients of the effective EMT components $\tau_t$ and $\tau^r$, respectively

\subsection{Limits}\label{ap-trans:3}	

The approximation of $e^h$ via Eq.~\eqref{fro-h}
\be
e^h\approx\sqrt{b^2+\frac{\xi}{x}}+d,
\ee
allows one to match the expansion of    Eq.~\eqref{hbarh} up to the $h_{12}$ term for all finite values of $r'_\sg$ (and, therefore, $\xi$ and $\Upsilon$).

The results of Section \ref{1-sol} set the constraint
\be
\frac{|r_\sg'|}{|r'_+|}=\frac{(1-w_1)\sqrt{\xi}}{2 \sqrt{\pi r_\sg^3}\Upsilon} .
\ee
Assuming $w_1=0$, Page's evaporation law and the static limit, $c_1\to w_1$, we find using the explicit expressions for $c_1$, $h_{12}$ and $w_1$ that
\be
h_{12}\simeq \frac{1}{4 \sqrt{\pi r^3_\sg}\Upsilon},
\ee
and thus
\be
d=\sqrt{\xi}h_{12}=\frac{1}{2}, \qquad  b^2=\frac{1}{4}.
\ee
Using this approximation for $e^h$ and $f\approx \alpha_{12}\sqrt{x}+x/r_\sg$, we obtain for the incoming null ray
\be
\frac{dt}{dr}=-\frac{1}{e^hf}\approx-\frac{2 r_\sg}{(\sqrt{x}+\alpha_{12} r_\sg)(\sqrt{x}+\sqrt{x+4\xi})}\ ,
\ee
that can be integrated in the closed form. For $r'_\sg\to 0$ the  propagation  time from $x_\mathrm{in}\sim x_*$ to $x_\mathrm{f}$ according to Bob changes from
$\Delta t\sim (x_\mathrm{in}-x_\mathrm{f})/|r_\sg'|$   to the logarithmic divergence $\Delta t\sim r_\sg \ln (x_\mathrm{in}/x_\mathrm{f})$.

\section{Separatrix}\label{a-sep}

Expanding Eq.~\eqref{YFsep} in the first order in $y=r-r_+$ results in
\begin{widetext}
\begin{align}
	y_\mathrm{sep}= \frac{2r_+(1-w_1)r'_+}{ 1+2w_1^2+\big(1+2 r_+(w_2-\chi_1)\big)r'_+ -2\big(1+(1-2\chi_1 r_+)r'_+\big)w_1-2r_+w'_1  }.
\end{align}
\end{widetext}		
Keeping only the leafing order terms in $|r'_+|\ll 1$ and $w_1\approx 0$ we obtain Eq.~\eqref{app:sep}.	

\section{Painlev\'{e}--Gullstrand coordinates} \label{ap-PG}
Among the families of coordinates that are regular across the horizon \cite{NV:06,VAC:11,MP:01} there are two types of the PG coordintes $(\bar t,r)$, where
\be
\bar t=t +F(t,r)
\ee
  is either the proper time of an infalling observer (with zero initial velocity at infinity) as the time coordinate, or chosen in such a way that the slices of constant $\bt$ have zero curvature. The two definitions are in general inequivalent, and it is the latter that is used in dynamical generalisations of the surface gravity. The flat slice condition leads the first-order linear  partial differential equation
  \be
   \pad_r F= \sqrt{\frac{C}{r}}\frac{e^{-h}}{f}(1+\pad_t F)=\eqdef \cA\pad_t\bt , \label{pde1}
  \ee
where the sign choice is determined by the agreement with the standard (ingoing) PG coordinates for the Schwarzschild black hole.

Subject to appropriate boundary conditions this equation has a unique solution. The metric in $(\bt,r)$ coordinates is then
\begin{align}
	ds^2= -\frac{e^{2h}}{(\pad_t \bt)^2}d\bt^2 +2 \frac{e^h}{\pad_t\bt}\sqrt{\frac{C}{r}}d\bt dr+dr^2+r^2d\Omega,  \label{pgg}
\end{align}
 For definiteness we choose $r_0>r_\sg(t)$ such that
 \be
 \bt(t,r_0)=t ,
 \ee
 as the initial condition.

In a slowly evolving case (where the evolution scale is set by $|r'_\sg|\ll1$), the PG coordinate is obtained iteratively. We set
\be
\bt=t+F_0(t,r)+F_1(t,r),
\ee
where $F_0$ satisfies $\pad_r F_0=\cA$ and thus
\be
F_0=\int_{r_0}^r\cA(t,\tilde r)d\tilde r.
\ee
Then, instead of looking for the exact characteristic curves of the equation, we use the slow evolution condition to approximate
\be
F_1\approx\int_{r_0}^r\cA(t,\tilde r)\pad_t F_0(t,\tilde r)d\tilde r.
\ee

Using the near-horizon metric of Eqs.~\eqref{Ctr} and \eqref{htr}, as well as expressions for the expansion coefficients in Appendix  \ref{ap-trans:2}	we get
\be
\cA=a_0+a_{12}\sqrt{\frac{x}{\xi}}+\cO(x) ,
\ee
where $a_0=1/|r'_\sg|$, and
\be
a_{12}= -\frac{2(1-w_1)}{8\pi r_\sg^2\Upsilon^2}-\frac{1}{2}.
\ee
It is further simplified if we assume $w_1=0$ and then Page's evaporation law that results in Eq.~\eqref{al2rpr}. Hence,
\be
a_{12}=\frac{1}{2}\left(1-\frac{\alpha^2}{r'_\sg{^2}}\right)\approx -\frac{1}{|r'_\sg|}.
\ee
Hence
\be
F_0=\frac{x-x_0}{|r'_\sg|}+\frac{2a_{12}}{3\sqrt{\xi}}\big(x^{3/2}-x_0^{3/2}\big)+\cO\big(x^2\big),
\ee
where $x_0=r_0-r_\sg$. Note that the magnitude of the second term is smaller than that of the first only if $x_0\lesssim \xi$, and it dominates for $x\gg \xi$. It is straightforward to obtain the leading contribution to $F_1$, but its expression is rather cumbersome. For our goal ($\pad\bt(t,r)/\pad t$ on the apparent horizon), we can retain only the first term in $F_0$. Then
\be
F_1\approx -\frac{(r_0-r)^2 r''_\sg}{2 (r'_\sg)^3},
\ee
and thus near the apparent horizon $F_1/F_0=(r_0-r)/r_\sg$. As a result the near-horizon fom of the  frozen metric in the PG coordinates is
\be
\bt=t+(r-r_0)/r_\sg+\cO\big(x^{3/2}).
\ee

	\end{document}